\documentclass[a4paper]{jpconf}
\usepackage{graphicx}
\usepackage{upgreek,xspace}
\usepackage{hyperref}
\def\textalpha{\ensuremath\upalpha}
\def\textbeta{\ensuremath\upbeta}
\def\textgamma{\ensuremath\upgamma}

\def\textmu{\ensuremath\upmu}
\begin{document}
\title{Status of the DEAP-3600 experiment}

\author{Marcin Ku\'zniak for the DEAP-3600 collaboration}

\address{AstroCeNT, Nicolaus Copernicus Astronomical Center of the Polish Academy of Sciences, Rektorska 4, 00-614 Warsaw, Poland}

\ead{mkuzniak@camk.edu.pl}

\begin{abstract}
DEAP-3600 is a single-phase liquid argon (LAr) dark matter detector, located 2~km underground at SNOLAB in Sudbury, Canada, which started taking data in 2016. The detector is sensitive to nuclear recoils induced by scattering of dark matter particles, which would cause emission of prompt scintillation light. DEAP-3600 demonstrated excellent performance, holds the leading WIMP exclusion among LAr detectors, and published several physics results. The WIMP sensitivity of the detector is currently limited by backgrounds induced by alpha activity at the LAr inlet, in a shadowed region of detector. The ongoing hardware upgrade aims at fixing that limitation and, in consequence, at reaching the full WIMP sensitivity. This paper summarizes the latest results from DEAP-3600 and details of the upgrade.
\end{abstract}

\section{Introduction}
The existence of dark matter is well established through astronomical observations, by a number of independent data sources and with complementary techniques. A new, as of yet undetected, type of particle is a candidate for dark matter, with Weakly Interacting Massive Particle (WIMP) being one of the favored scenarios. Direct detection of such particles is attempted by looking for evidence of their interactions with ordinary matter. A number of ultra-low background detectors used for this purpose are located in deep underground laboratories, in order to evade backgrounds induced by cosmic rays. For a detailed discussion of the motivation and status of direct experimental searches, see the recent review~\cite{appec}.

\section{The DEAP-3600 detector}
DEAP-3600 was built to register scintillation light induced by elastic scattering of WIMPs on argon nuclei within a 3.3-tonne liquid argon (LAr) target, contained in a spherical acrylic vessel (AV)~\cite{deap}. The AV is coated on the inside with tetraphenyl butadiene (TPB) wavelength shifter (WLS), so that the LAr scintillation light, peaked at 128~nm i.e. in vacuum ultraviolet (VUV), is converted to visible for efficient detection with 255 photomultiplier tubes (PMTs) surrounding the AV. 

In order to reduce backgrounds from natural radioactivity, mainly neutrons and \textalpha\ particles, the detector is: (1) constructed from carefully selected radiopure materials, (2) shielded from ambient radioactivity by a water tank instrumented with PMTs, which also serves as a Cherenkov muon veto, and (3) located underground at SNOLAB, where the muon flux is low.

\textgamma-induced electron recoil (ER) events are rejected with pulse-shape discrimination (PSD), which is exceptionally powerful in LAr because of very significant difference between lifetimes of the excited LAr triplet ($\sim$1.5~\textmu s) and singlet ($\sim$6~ns) states, which are preferentially excited by ER and nuclear recoils, respectively. This property of LAr, together with its high scintillation yield, ease of purification and high natural abundance, makes it an ideal and scalable target for large dark matter detectors.

In its second run the DEAP-3600 collected data from November 2016 until March 2020, when it was emptied of LAr in order to prepare it for a COVID-19 related period of reduced maintenance, as well as for a planned hardware upgrade. The physics data collection remains 80\% blind in the WIMP search region of interest (ROI) since January 2018.

\section{Recent physics highlights}
Since the commissioning phase in 2016, DEAP-3600 holds the leading spin-independent WIMP-nucleon scattering cross section upper limit measured in a LAr target~\cite{PRL,PRD}. The latest null result from 231 live days (exposure of 758 tonne-days) was 3.9$\times$10$^{-45}$~cm$^2$ (1.5$\times$10$^{-44}$ cm$^2$) for a 100~GeV/c$^2$ (1~TeV/c$^2$) WIMP mass at 90\% C.\,L., assuming the Standard Halo Model (SHM) and isospin parity conservation.

\begin{figure}[htb]
\centering
\includegraphics[width=0.43\columnwidth]{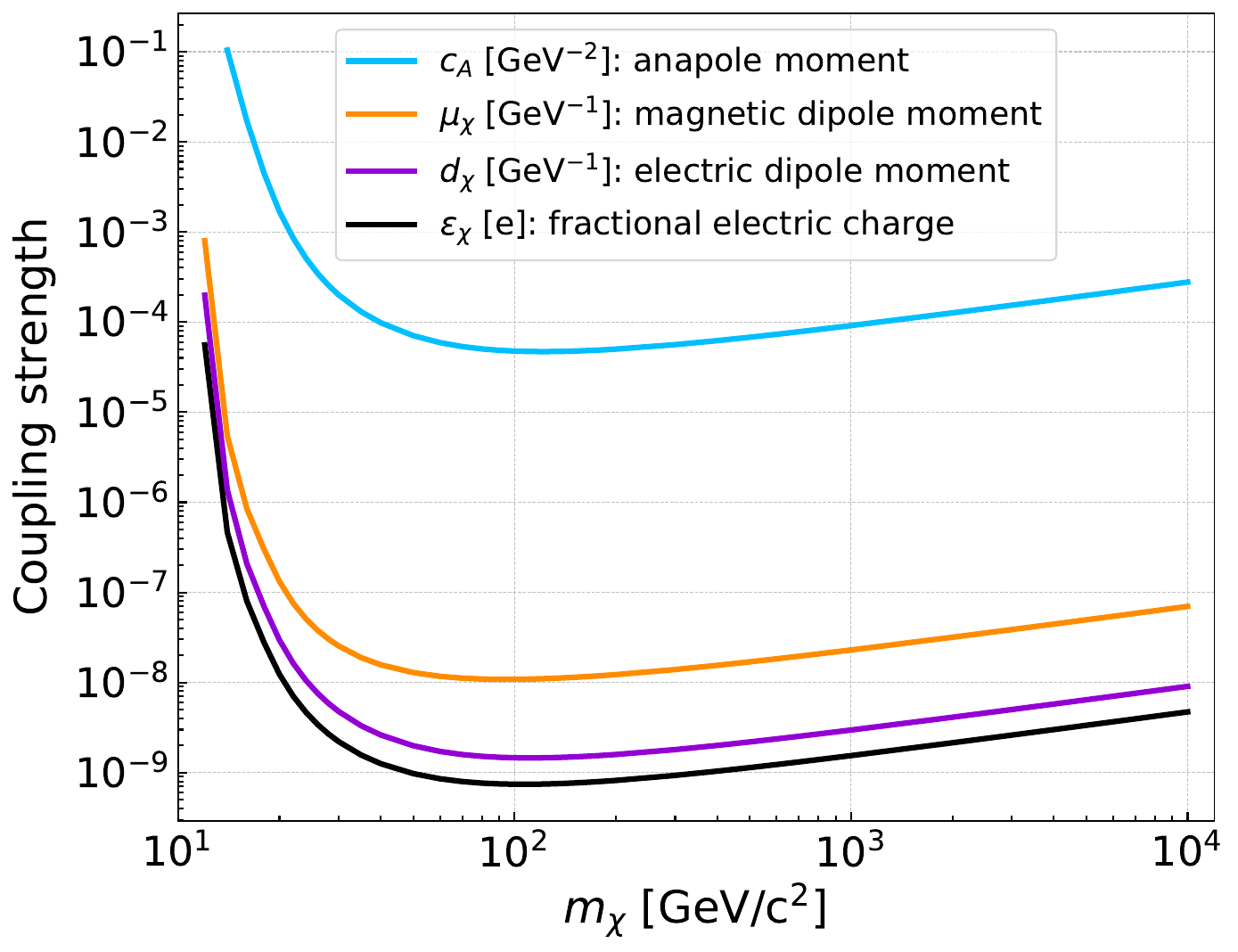}\includegraphics[width=0.438\columnwidth]{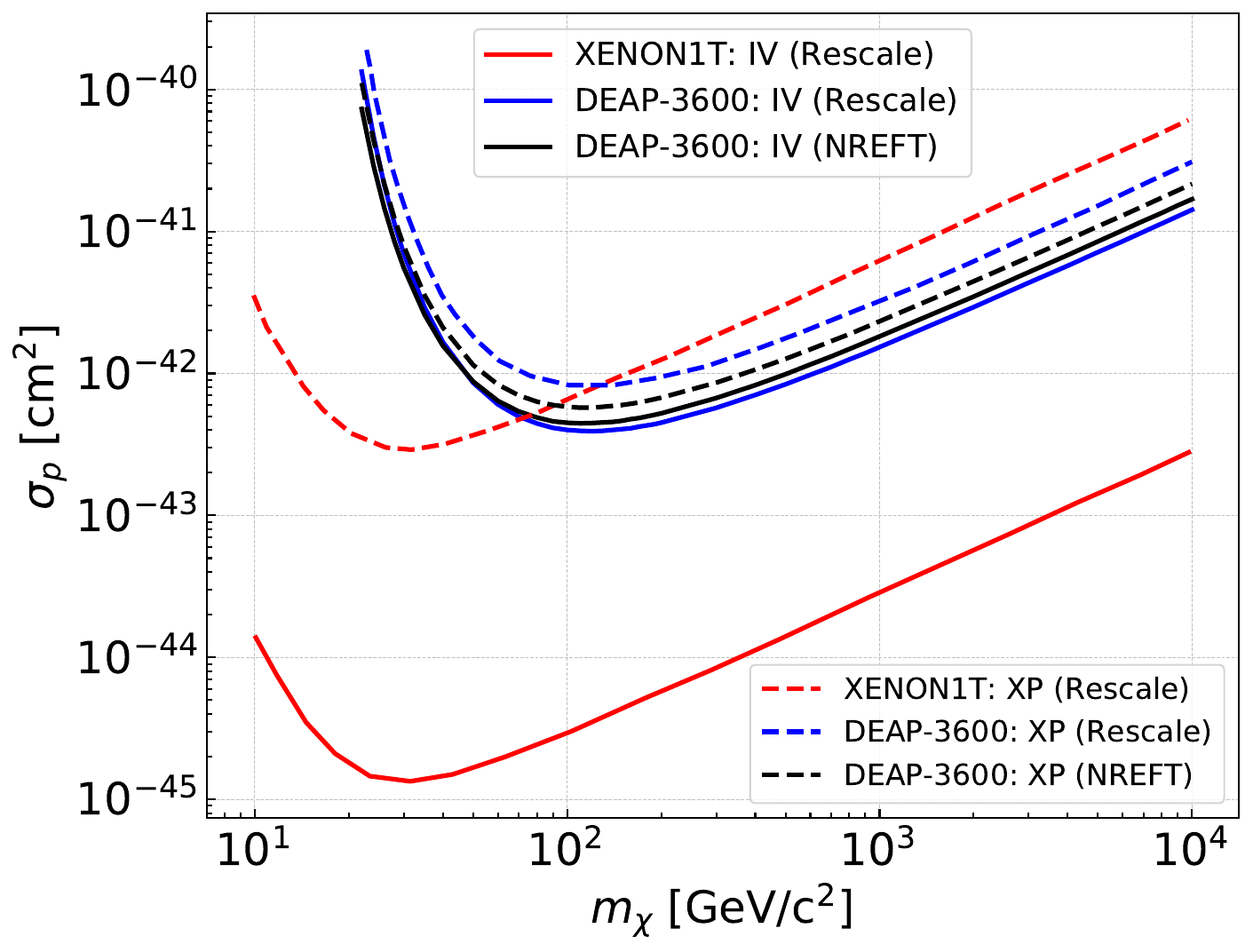}
\caption{(Left) Exclusion curves on the coupling strength of photon-mediated interactions: anapole, magnetic dipole, electric dipole and millicharged DM using the SHM. (Right) Constraints on the O$_1$ interaction from XENON1T and DEAP-3600, for IV (isovector; solid) and XP (xenonphobic; dashed) scenarios from non-relativistic EFT (NREFT). For more details see Ref.~\cite{EFT}.}
\label{fig:eft}
\end{figure}

Recently, this result has been re-interpreted from the standpoint of Effective Field Theory (EFT) and astrophysical uncertainties on the dark matter halo substructure~\cite{EFT}. In addition to studying the effects of stellar streams and debris flows on the DM velocity spectrum and the resulting WIMP exclusions, the DEAP-3600 result has been also translated to upper limits on coupling strength of different non-relativistic EFT (NREFT) contact interaction operators, isovector couplings, as well as millicharge, magnetic dipole, electric dipole and anapole WIMP-nucleon interactions, see Fig.~\ref{fig:eft}(left). Isospin-violating scenarios have also been studied, including the so-called xenonphobic (XP) coupling scenario, where WIMP-proton and WIMP-neutron interactions are such that the DEAP-3600 limits are world leading at high WIMP mass, see Fig.~\ref{fig:eft}(right).

\begin{figure}[htb]
\centering
\includegraphics[width=0.435\columnwidth,trim=0 13cm 0 0,clip=true]{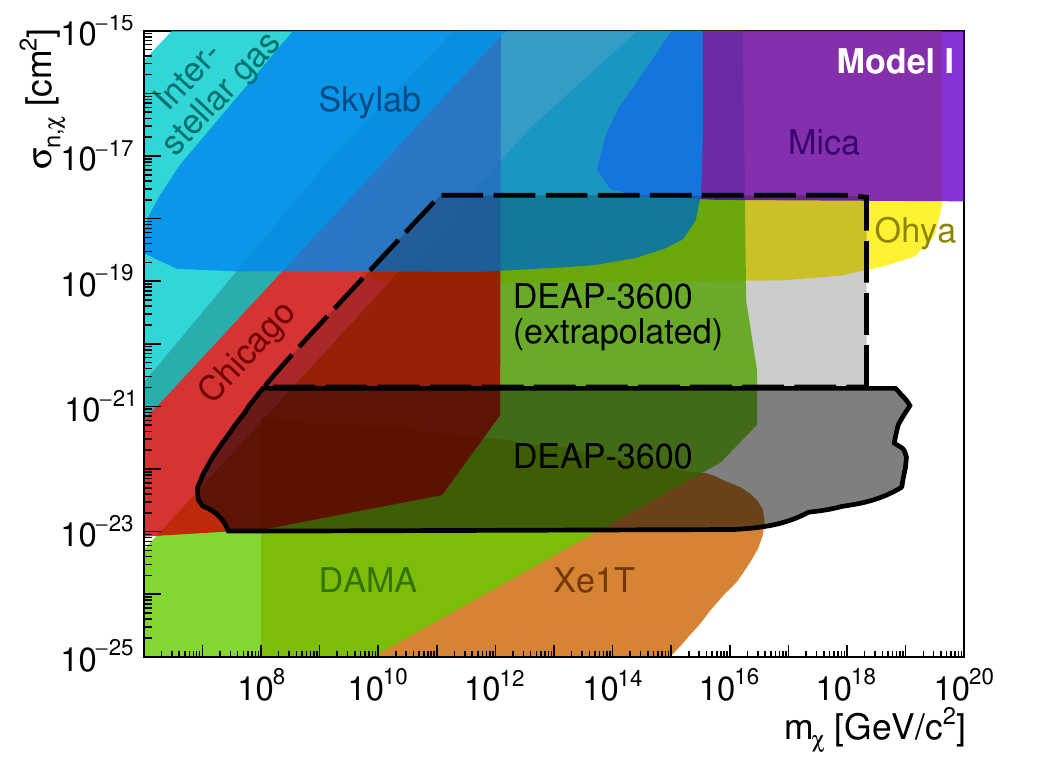}\includegraphics[width=0.435\columnwidth,trim=0 0 0 13cm 0,clip=true]{hig.pdf}
\caption{Exclusions at 90\% C.\,L. set in the Planck scale mass dark matter search, compared with results from other experiments, for two theoretical models: (Left) opaque dark
matter candidate, relevant for composite DM models, (Right) assuming the coherent scattering cross section, scaling with A$^4$. For more details, see Ref.~\cite{MIMP}.}
\label{fig:mimp}
\end{figure}

The full available dataset, 813 live days, was employed for a blind search of Planck-scale mass dark matter probing so far unexplored parameter space~\cite{MIMP}. Multi-site scatter in the detector was the key signal signature for such super-heavy particles with high scattering cross section. The null search result permitted setting new leading limits, see Fig.~\ref{fig:mimp}.

\section{Backgrounds}
The DEAP collaboration has pioneered the use of pulse-shape discrimination in LAr for dark matter detection, essential because of the presence of 0.95~Bq/kg activity of cosmogenically produced $^{39}$Ar in atmospheric Ar. More recently, a complete model of the LAr scintillation event pulse shape was developed, taking into account the LAr time constants, WLS re-emission time profile and PMT-related noise contributions from, primarily, charge resolution, dark noise rate and afterpulsing~\cite{pulse}. That model, combined with an in-situ calibration of noise characteristics of each individual PMT and a Bayesian single photoelectron counting algorithm allowed for an unprecedented PSD performance, resulting in a fraction of \textbeta\ events misidentified as NR of 10$^{-9}$ at the 16~keV threshold~\cite{PSD}.
Using demonstrated reduction in $^{39}$Ar from underground argon, this is projected to $<$0.2 events in 3000~tonne-years in a large single-phase detector (such as ARGO~\cite{argo}).

The ER background model, including \textbeta\ and \textgamma\ from internal and external radioactivity, informed by ex-situ assay data, was shown to be close to the observed spectrum. A multi-component fit, which achieved good agreement with data over nine orders of magnitude in event rate, allowed to measure each contribution, confirm assumptions informing the neutron rate expectation, and, as a by-product, provide a precise measurement of the $^{42}$Ar specific activity in the atmosphere~\cite{Ar}.

The neutron background is dominated by (\textalpha,~n) reaction in the detector materials, mainly the PMT glass. The neutron rate is estimated in-situ using a delayed coincidence analysis searching for neutron capture gammas following the NR, resulting in 0.10$^{+0.10}_{-0.09}$ expected number of neutron background events in the last WIMP search~\cite{PRD}.

\textalpha\ activity itself, from Rn-borne $^{210}$Pb/$^{210}$Po present in the surface layer of the AV since the detector construction, is efficiently removed with position reconstruction, leading to an expectation of $<$0.08~background events.
DEAP-3600 has also demonstrated the lowest so far among noble liquid detectors rate of suspended Rn events from constant emanation, approx.~0.15~\textmu Bq/kg (4~nBq/kg) for $^{222}$Rn ($^{220}$Rn)~\cite{PRD}.

The above demonstrates excellent control over conventional types of backgrounds, achieved with a single-phase detector, which makes it an attractive technology for scale-up. The remaining special two classes of \textalpha\ backgrounds, will be mitigated with the ongoing hardware upgrade.

The main limiting background for the previous DEAP-3600 result was the \textalpha\ activity in the LAr inlet on top of the detector, the so-called neck. While filled with gaseous Ar, it contains enough LAr film on the surfaces and/or mist for \textalpha's to induce prompt scintillation light, which is shadowed by the acrylic flowguides, resulting in misreconstructed events with apparent low energy. A detailed Monte Carlo model of such events has been developed, showing excellent agreement with the data and allowing for reliable extrapolations to the ROI. While a number of topological features can be used in the analysis to remove such events, it is only achieved at a cost of significantly reduced signal acceptance~\cite{PRD}.

Another issue is the evidence of \textalpha\ activity in the dust particulates present in the LAr volume. This results in events with \textalpha\ energy degraded in the particulate material, and a continuum of events extending to low energies, dependent on the particulate size distribution and material. A power law distribution describes the data very well at intermediate energies, however, a reliable extrapolation to the WIMP ROI will be made once the systematic uncertainty is constrained by characterization of the dust extracted from the detector. Similarly to neck events, topological features of dust events are explored with multivariate analysis, due to the expected effect of partial shadowing of the scintillation light by dust particulates.

\section{Hardware upgrade}
While analysis tools are being refined for better rejection of \textalpha\ backgrounds in the dataset already collected, a hardware upgrade of the detector aimed at eliminating the issue, in addition to the necessary maintenance on cryogenic systems, is currently in progress. For the neck backgrounds two independent preventive measures have been proposed. 

Firstly, a new set of acrylic flowguides has been machined and sanded in low-Rn atmosphere, and subsequently coated with a WLS with a distinctly slow re-emission time constant. In consequence, the VUV photons from \textalpha\ scintillation directly reaching the inside of the detector through the opening in the neck, will be dominated by delayed visible photons from the slow WLS, enabling efficient PSD, and without significant effects on the WIMP signal acceptance. The collaboration has optimized the composition and fully characterized performance of such coatings, converging on pyrene-doped polystyrene, which exhibits the light yield of 36-46\% of TPB and the fluorescence lifetime of 279(14)~ns, confirmed at cryogenic conditions and VUV excitation wavelengths, as well as successfully tested these for mechanical robustness~\cite{pyrene,David,Hicham}. Based on Monte Carlo simulations, a leakage fraction of $<$1.2$\times$10$^{-5}$ is expected with an optimized PSD discriminant, more than sufficient to remove such events from the WIMP search ROI.

Secondly, an alternate cooling and LAr delivery system has been designed and fabricated in order to: (1) allow for extraction and filtration of dust particulates from the detector and (2) increase the temperature of the neck surfaces, preventing formation of the layer of liquid or mist in that region, which additionally prevents occurrence of prompt, WIMP signal-like, LAr scintillation induced by \textalpha's. A stainless steel vacuum-jacketed pipe deployed all the way to the bottom of the AV will be initially used for extraction of LAr. Then, lifted to the top of the AV and connected to an externally located argon liquefier, it will allow for injection of LAr into the detector and operation with the previously-used neck cooling coil turned off, resulting in the temperature of the neck increased above the Ar boiling point.

Installation and commissioning are planned for Winter and Spring, followed by filling the detector and data taking, starting from Summer 2022.

\section{Summary and outlook}
The DEAP-3600 detector has demonstrated the potential of single-phase technology showing excellent performance and control over the key \textbeta/\textgamma, neutron and surface \textalpha\ backgrounds. The sensitivity of the detector is currently limited by neck and dust particulate \textalpha\ backgrounds, which are being addressed in the ongoing detector upgrade with slow WLS coatings in the neck region and a particulate extraction/filtration system; the upgraded detector will be able to reach its design sensitivity. In parallel, multivariate analysis techniques are being employed to improve the sensitivity of the WIMP search using the dataset already collected; which will be informed by the characterization of the dust sample extracted from the detector.

Recent analysis highlights include the NREFT-based reinterpretation of the last WIMP exclusion, as well as leading sensitivity constraints on the Planck scale mass dark matter.

The collaboration is working on a number of other physics analyses, including the precision measurement of specific activity of $^{39}$Ar in atmospheric argon and the first measurement of the $^8$B solar neutrino absorption.

The detector is expected to resume data taking in Summer 2022 and will proceed with the physics run, as the currently most sensitive LAr detector. In addition to probing new physics, DEAP-3600, as a single-phase detector, is an important test-bed that will inform the scale up to the future ultimate LAr experiment, capable of probing the WIMP paramter space down to the neutrino floor~\cite{argo}.
\ack
We thank the Natural Sciences and Engineering Research Council of Canada,
the Canadian Foundation for Innovation (CFI),
the Ontario Ministry of Research and Innovation (MRI), 
and Alberta Advanced Education and Technology (ASRIP),
Queen's University,
the University of Alberta,
Carleton University,
the Canada First Research Excellence Fund,
the Arthur B.~McDonald Canadian Astroparticle Research Institute,
DGAPA-UNAM (PAPIIT No.~IN108020) and Consejo Nacional de Ciencia y Tecnolog\'ia (CONACyT, Mexico, Grant A1-S-8960),
the European Research Council Project (ERC StG 279980),
the UK Science and Technology Facilities Council (STFC) (ST/K002570/1 and ST/R002908/1), 
the Russian Science Foundation (Grant No.~16-12-10369), 
the Leverhulme Trust (ECF-20130496),
the Spanish Ministry of Science and Innovation (PID2019-109374GB-I00), 
the International Research Agenda Programme AstroCeNT (MAB/2018/7)
funded by the Foundation for Polish Science (FNP) from the European Regional Development Fund, and from the European Union’s Horizon~2020 research and innovation programme (DarkWave, Grant No. 952480).
Studentship support from
the Rutherford Appleton Laboratory Particle Physics Division,
STFC and SEPNet PhD is acknowledged.
We thank SNOLAB and its staff for support through underground space, logistical, and technical services.
SNOLAB operations are supported by the CFI
and Province of Ontario MRI,
with underground access provided by Vale at the Creighton mine site.
We thank Vale for their continuing support, including the work of shipping the acrylic vessel underground.
We gratefully acknowledge the support of Compute Canada,
Calcul Qu\'ebec,
the Centre for Advanced Computing at Queen's University,
and the Computation Centre for Particle and Astrophysics (C2PAP) at the Leibniz Supercomputer Centre (LRZ)
for providing the computing resources required to undertake this work.

\end{document}